\documentclass[journal]{vgtc}         

\ifpdf
  \pdfoutput=1\relax                   
  \usepackage{graphicx}                
  \DeclareGraphicsExtensions{.pdf,.png,.jpg,.jpeg} 
\else
  \usepackage{graphicx}                
  \DeclareGraphicsExtensions{.eps}     
\fi%

\graphicspath{{figures/}{pictures/}{images/}{./}} 

\usepackage{microtype}                 
\PassOptionsToPackage{warn}{textcomp}  
\usepackage{textcomp}                  
\usepackage{mathptmx}                  
\usepackage{times}                     
\usepackage{cite}                      
\usepackage{tabu}                      
\usepackage{booktabs}                  

\usepackage{enumitem}
\usepackage{float}
\usepackage{soul}

\makeatletter
 \def\SOUL@hlpreamble{%
 \setul{}{2ex}
 \let\SOUL@stcolor\SOUL@hlcolor
 \SOUL@stpreamble
 }
\makeatother

\onlineid{0}

\vgtccategory{Research}
\vgtcpapertype{please specify}

\title{VizCommender: Computing Text-Based Similarity in\\Visualization Repositories for Content-Based Recommendations}

\author{Michael Oppermann, Robert Kincaid \textit{Senior Member, IEEE}, and Tamara Munzner \textit{Senior Member, IEEE}}
\authorfooter{
\item 
 Michael Oppermann is with Tableau Research and the University of British Columbia. E-mail: opperman@cs.ubc.ca.
\item
 Robert Kincaid is with Tableau Research (retired). E-mail: robert.kincaid@startmail.com.
\item Tamara Munzner is with the University of British Columbia. E-mail: tmm@cs.ubc.ca
}

\shortauthortitle{Biv \MakeLowercase{\textit{et al.}}: Global Illumination for Fun and Profit}

\abstract{
Cloud-based visualization services have made visual analytics accessible to a much wider audience than ever before. Systems such as Tableau have started to amass increasingly large repositories of analytical knowledge in the form of interactive visualization workbooks. When shared, these collections can form a visual analytic knowledge base. However, as the size of a collection increases, so does the difficulty in finding relevant information. Content-based recommendation (CBR) systems could help analysts in finding and managing workbooks relevant to their interests. Toward this goal, we focus on text-based content that is representative of the subject matter of visualizations rather than the visual encodings and style. We discuss the challenges associated with creating a CBR based on visualization specifications and explore more concretely how to implement the relevance measures required using Tableau workbook specifications as the source of content data. We also demonstrate what information can be extracted from these visualization specifications and how various natural language processing techniques can be used to compute similarity between workbooks as one way to measure relevance. We report on a crowd-sourced user study to determine if our similarity measure mimics human judgement. Finally, we choose latent Dirichlet allocation (LDA) as a specific model and instantiate it in a proof-of-concept recommender tool to demonstrate the basic function of our similarity measure.
} 


\CCScatlist{ 
 \CCScat{K.6.1}{Management of Computing and Information Systems}%
{Project and People Management}{Life Cycle};
 \CCScat{K.7.m}{The Computing Profession}{Miscellaneous}{Ethics}
}

\teaser{
 \centering
 \includegraphics[width=\linewidth]{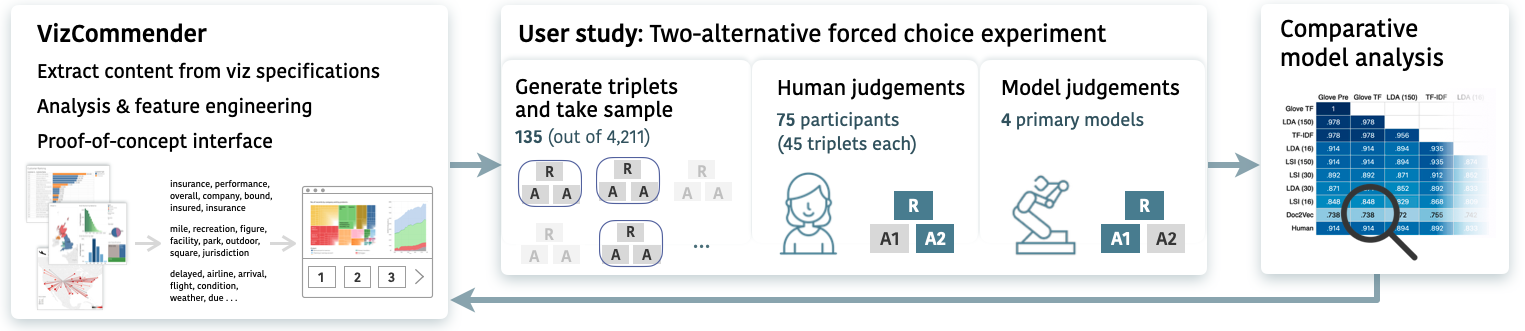}
    \vspace{-18pt}
    \caption{Outline of our process: (1) Extract and analyze textual content from viz specifications belonging to two VizRepos. Investigate appropriate content-based recommendation models with varying input features and implement initial prototypes to facilitate discussions with collaborators; (2)~Crowdsourced study: Sample viz triplets in a semi-automated process, collect human judgements about the semantic text similarity, and run the same experiment with different NLP models; (3) Compare agreement between human judgements and model predictions to assess model appropriateness. Implement LDA-based similarity measure in proof-of-concept pipeline.}
     \label{fig:teaser}
}

\vgtcinsertpkg

\begin{document}


\firstsection{Introduction}

\maketitle
As information visualization and visual analytics have matured, cloud-based visualization services have emerged. Tableau, Microsoft Power BI, Looker, and Google Data Studio are a few such examples. In addition to enabling the sharing of individual one-off visualizations, users collaboratively build shared repositories that serve as visual analytic knowledge bases. 
By providing large-scale community- and organization-based collaboration services~\cite{viegas2007manyEyes}, massive repositories of visual data representations are created over time, referred to as \mbox{\textit{VizRepos}} in the remainder of this paper. The primary artifacts in VizRepos are \textit{visualization workbooks} (or reports) that bundle a set of visualizations or dashboards regarding a specific task or data source. VizRepos exhibit the same issues of ad hoc organization, inconsistent metadata, and scale that are common in the larger context of the web itself and similar large-scale item collections such as streaming services~\cite{zhou2016youtube}, web-based shopping portals~\cite{mauge2012structuringEcommerce}, digital libraries~\cite{witten2009digitalLibrary}, or data lakes that transform into data swamps~\cite{brackenbury2018drainingDataSwamp,hai2016constance}. Thus, cloud-based visualization systems are also encountering an increasing need for efficient discovery of relevant visualization artifacts, much like these more well-known platforms.

As with all such repositories, many possible approaches can be employed to find relevant information. Indexed  searching~\cite{white2009exploratorySearch}, faceted navigation~\cite{yee2003facetedSearch}, or advanced filter options are common methods. In addition to these active querying methods, recommender systems~\cite{aggarwal2016recSys} are increasingly used to passively assist users by surfacing relevant content, depending on the individual context.

A major paradigm for recommender systems is content-based filtering that incorporates available content features into a domain-specific model rather than looking at generic interactions between users and items, or ratings. \textit{Content-based recommendation (CBR) systems} exist for a wide range of data types, such as songs~\cite{vall2019playlistRecSys}, videos~\cite{deldjoo2016videoRecSys}, artworks~\cite{messina2019artworkRecSys}, and news articles~\cite{kompan2010newsRecSys}. For VizRepos, however, content-based recommendation has not been explicitly addressed. Content features are highly subjective and imprecise in nature~\cite{zenebe2009fuzzySetsRecSys}, and thus CBR systems for new kinds of data require custom feature engineering building upon extensive domain expertise in order to make meaningful comparisons and rankings.

VizRepos, and more specifically visualization workbooks, are based on a set of static \textit{visualization specifications} that contain information about the data source and any data transformations, how the data is presented to the user through a visual encoding, and how users can interact with the visualization. Visualization specifications have an unusual combination of characteristics such as sparse and often fragmented text, hierarchical semi-structured content, both visual and semantic features, and domain-specific expressions. Our goal is to extract suitable content features to perform similarity comparisons between visualization specifications that can ultimately provide the basis of content-based recommendations for VizRepos. This process is driven by two key questions: (1) Which content features are most informative for comparing visualization specifications? and (2) What techniques can we use for comparing and ranking visualization specifications?

In this paper, we describe our work towards developing a text-based similarity measure suitable for content-based recommendations in VizRepos. 
We focus the notion of similarity on the subject matter of the visualizations rather than the specific visual encoding, with the assumption that similar topics are of high relevance. Initial investigations indicated that chart types and visual styles are of peripheral interest in the information seeking process.

We examine the performance of multiple existing natural language processing models within our proposed similarity model, and find that many of them yield robust results that align with human judgements, despite the challenging characteristics of the textual data that can be extracted from visualization workbooks.

Our contributions are:

\begin{itemize}[parsep=0pt,topsep=0pt]
\item Identification and characterization of unique challenges in designing content-based recommendations for visualization repositories.

\item Design and implementation of the VizCommender proof-of-concept pipeline: Textual feature extraction, similarity measures for comparing and ranking visualizations using natural language processing (NLP) techniques, and a user interface to demonstrate and provide insight into the measure's utility for visualization recommendations.

\item Analysis of four applicable NLP similarity measures by examining both computational comparisons and a user study assessing alignment with human judgements of similarity.

\end{itemize}

In our work, we align with the terminology in the recommender system literature~\cite{aggarwal2016recSys} where recommendations are proposed for preexisting items of content, namely previously designed visualizations stored in VizRepos as specifications. In contrast, the term \textit{visualization recommendation} is often used in visualization research quite differently, to refer to recommending a new \textit{visual encoding} for a given dataset~\cite{vartak2017towardsVisRec}, for example with Tableau's ShowMe~\cite{mackinlay2007showMe} or Draco~\cite{moritz2018draco}. In this paper, we do not focus on generating new visual encodings.

\section{Background}

We briefly discuss general approaches to recommender systems and elucidate our industry collaboration, followed by a description of our target users and a data characterization.

\subsection{Recommendation Paradigms}

Many different types of recommender systems have been proposed and are widely disseminated for many domains~\cite{aggarwal2016recSys}. We differentiate between the two major recommendation paradigms: \textit{collaborative filtering} (CFR) and \textit{content-based filtering} (CBR).

CFR analyzes the interaction between users and items and derives recommendations from patterns of similar user behavior or ratings. This approach requires no domain knowledge, allows fast computation, and provides diverse and often serendipitous recommendations. However, CFR suffers from the so-called \textit{cold start} problem when new items or new users emerge and the system does not have sufficient information to make recommendations~\cite{lam2008coldStart}.

CBR is the focus of this paper and is based on finding relevant items based on their actual content. Since this method is based on inherent properties of the items it does not suffer from the cold start problem of collaborative filtering and can frequently be more accurate depending on the recommendation needs. In contrast to CFR, this approach also allows one to identify \textit{near-duplicate items} which are highly relevant in some cases but need to be excluded in other cases~\cite{vliegendhart2012userPerceptionNearDuplicate}. Approaches to CBR systems have been widely investigated; the challenge is to selecting what content features should be considered and how they are to be compared, which is known to be highly domain-dependent~\cite{lops2011content}.

Ideally, both CFR and CBR approaches are combined into a hybrid system to alleviate the issues of the other. CBR can make useful recommendations for new systems while user behavior is monitored. Once sufficient information is collected that CFR becomes useful, then serendipity and diversity can be increased in the recommendations.


\subsection{Project Context}


This work was conducted at Tableau Research. This environment provided unique access to multiple domain experts in visualization research, machine learning, and most importantly customer insight into typical use cases and pain points related to managing large VizRepos and finding relevant information. We had over eight months of continuous engagement with product managers, user experience designers, and multiple machine learning engineers in the Tableau Recommender Systems Group. The overall approach, feature engineering, and algorithm choices were informed, encouraged, and reviewed by these experts on an ongoing, iterative basis.

Tableau has already deployed CFR systems to present users with  relevant data sources in their data preparation workflow and to provide personalized viz recommendations in a dedicated section on Tableau Online. The latter application is most relevant for this work and motivated our investigation of CBR approaches because the standard user-item CFR model lacks precision and faces cold start issues. In particular, when viewing a specific viz workbook, the corresponding recommendations should be related to the reference. 
Ideally, advantages of both CFR and CBR are leveraged by combining them in a hybrid model. Although a hybrid system is beyond the scope of this paper we hope to inform the development of future hybrid systems with our work.

This collaboration allowed us to get access to two Tableau VizRepos with thousands of hand-crafted visualizations: (1) A sample of 29,521 workbooks (147,241 visualizations) that are publicly shared on Tableau Public, a community platform with more than two million workbooks; (2) An internal Tableau corporate repository with 3,661 workbooks (54,715 visualizations). Since Tableau Public can be used by anyone for free, that community-generated content is inherently messy and contains many mixed-language and incomplete workbooks, as described in~\autoref{sec:data-characterization}. We knew from our collaborators that the sprawling Tableau Public dataset has very different usage patterns than repositories used in more typical enterprise settings, so we also used an internal Tableau company repo in our experimentation and when obtaining expert feedback on versions of our proof-of-concept recommender. All data used in the crowd-sourced experiment and the examples shown in this paper are from the public dataset, for data security reasons.

After excluding workbooks with no visual encodings and non-English content, we obtained one Tableau Public VizRepo with 18,820 workbooks and one Tableau corporate VizRepo with 3,424 workbooks.



\begin{figure*}[ht]
 \centering
 \includegraphics[width=\linewidth]{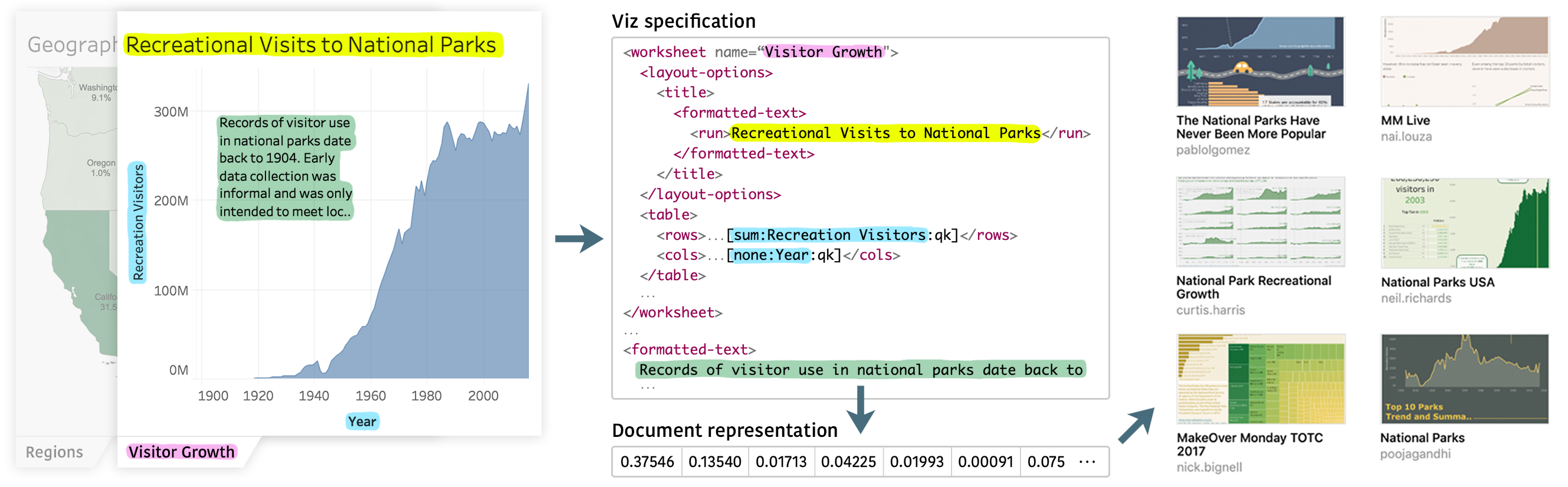}
 \vspace{-16pt}
 \caption{Simplified example feature extraction from a Tableau workbook. To illustrate typical features, a highly abbreviated example of the workbook XML is shown in the middle. Highlighted text color indicates the corresponding features that are converted into a numeric vector representation and used to compute text-based similarity. Similar types of text gets extracted from the remaining views and dashboards of the workbook.}
 \label{fig:vis-specification}
 \vspace{-12pt}
\end{figure*}

\subsection{Process}

We followed an iterative process, shown in~\autoref{fig:teaser}, that began by examining data from the provided VizRepos and by identifying and synthesizing use cases and recommendation challenges. In parallel to the investigation of appropriate models, we implemented a prototype user interface (\autoref{sec:proof-of-concept}) to analyze model outputs in a CBR context directly. The goal of VizCommender usage was to resemble the browsing experience for other VizRepos sufficiently to facilitate discussions and dialogue with project stakeholders, comparable to a technology probe~\cite{hutchinson2003technologyProbes}. For example, we investigated the use of visual versus topic features, and early demo sessions revealed that visual encodings are not as important as anticipated, as we describe in further detail in~\autoref{sec:leave-out-visual}.
The first analysis phase provided sufficient direction and informed the design of the crowdsourced study to systematically compare the alignment between human judgements and predictions of various NLP models (see \autoref{sec:study}).
The insights gained from informal model experiments, the study, and discussions with our collaborators were incorporated into the prototype which was continuously refined.

\subsection{Users and Tasks}

We began our project by understanding target users and tasks associated with a VizRepo recommender system. This process was informed by in-depth discussions with our collaborators, who have direct access to end-users and insight into their needs and pain points. The Recommender Systems Group also conducted a user study with a wizard-of-oz recommendation tool to investigate user needs.

The target users are \textit{explorers} who look up their own visualizations or browse a VizRepo for relevant work
that has been created by other users from the same organization or community. Our investigation revealed that information seeking or foraging is the core task when browsing VizRepos in enterprise settings. Information seeking is a well-known visual analytics task~\cite{shneiderman1996informationSeeking,choo2014visIRR,beck2015surVis}. In the context of CBR's this task is highly focused on finding informative workbooks around some specific analytical question in the user's mind. For instance, if the analyst inspects a specific workbook, additional relevant workbooks based on its topic or data should be automatically provided in-place.

Another task that is particularly focused on content features is version detection. This task supports quickly grouping of duplicate and near-duplicate items for users to compare different versions of workbooks and to maintain the organization of a repository.

One of the main insights of our requirements elicitation process was that a single set of recommendations is not adequate to capture user intent and the different tasks that should be facilitated by a recommender system. Our proposed recommendation facets are described in~\autoref{sec:proof-of-concept}.

We identified additional tasks that are not the focus of this paper. Recommender-assisted authoring is an interesting future possibility where recommendations are presented in real time while a user is authoring a new workbook. Such recommendations could be used for design inspiration, or one might even discover that a suitable workbook already exists in the repository. A more casual task is topic browsing. In this case a user may be simply browsing through the repository without any specific goal in mind other than to look for interesting workbooks and visualizations. Success in this task requires a more lax sense of relevance in order to surface serendipitous recommendations that venture further away from workbooks in the user's profile.


\subsection{Visualization Workbooks}

We use \textit{Tableau workbook specifications} as our source of visualization content, since we had ready access to large repositories of this data type. Bundling visualizations and data sources into workbooks or reports is a common practice between visualization services, such as Looker or Power BI, so our findings are generalizable beyond Tableau. For this paper, we will refer to Tableau workbooks as workbooks.

Workbooks are stored as XML documents describing all necessary components that are required for loading and displaying the fully-interactive visualizations to users. \autoref{fig:vis-specification} shows an example workbook and the corresponding XML excerpt.
A workbook typically contains linkages to one or more tabular data sources that are either static datasets, such as CSV data files, or live data streams. The viz specification includes the column names of all data attributes and any data transformations that have been applied, but not the raw data itself.

Individual visualizations referred to as \textit{views} are authored within containers called \textit{sheets} and can point to one or more data sources.
A sheet can alternatively represent a \textit{dashboard} which consists of references to individual sheets and additional layout information to result in a collection of related and possibly coordinated views.
A workbook could contain many hundred instances of views and dashboards; in practice, the majority of the workbooks that we analyzed have less than a dozen sheets, and about one third contain only a single view.

A complex set of XML tags are used to define the specifications for each of these objects. At a high level, one can consider a workbook specification to be structured very similarly to the typical XML-based spreadsheet which consists of multiple sheets, each with various data and style elements, and potentially with cross-references between sheets. The specification of a single view is comparable to the JSON-based view specification in Vega-Lite~\cite{satyanarayan2016vegaLite}.


\begin{figure*}
\begin{minipage}[c]{\columnwidth}
\includegraphics[width=\linewidth]{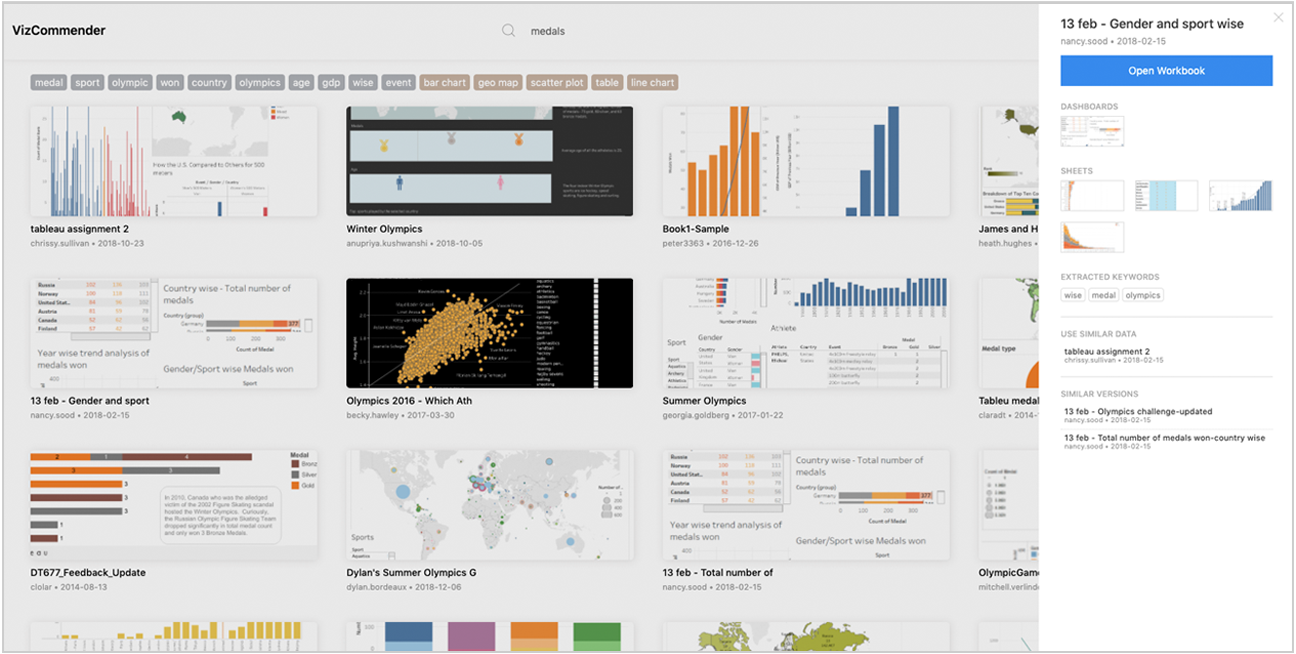}
 \caption{VizCommender interface that allows users to browse through a VizRepo. Workbook thumbnails are arranged in a grid view. Users can search for content or further drill down by selecting one of the tags at the top. The quick view sidebar on the right provides further details including recommendations when a workbook is selected.}
 \label{fig:explorer-overview}
\end{minipage}
\hfill
\begin{minipage}[c]{\columnwidth}
\includegraphics[width=\linewidth]{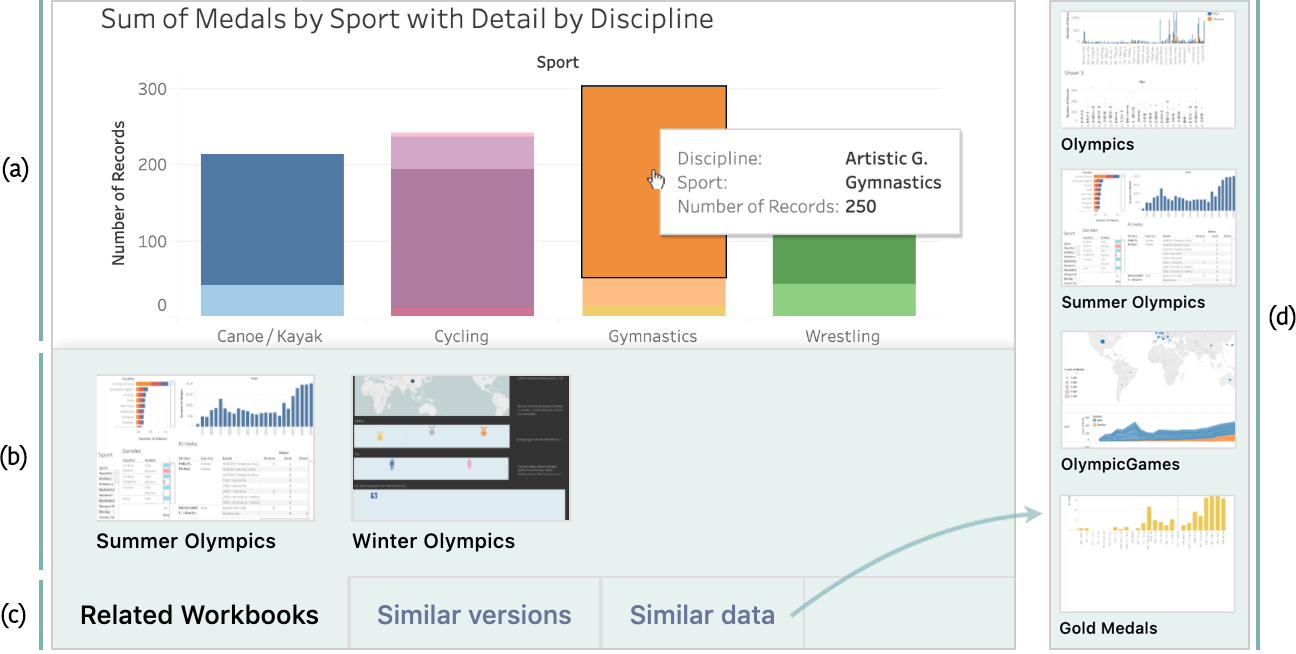}
 \caption{Interface detail view with recommendations. (a) Interactive Tableau workbook; (b) Expanded recommendation panel at the bottom of the screen showing related workbooks; (c) Tab navigation to switch between different recommendation types; (d) Alternative recommendation panel showing workbooks that use \textit{similar data}.}
 \label{fig:explorer-detail}
\end{minipage}%
 \vspace{-12pt}
\end{figure*}

\section{Related Work}

In this section we discuss related work on visualization recommendation, semantic text similarity models, as well as evaluation methods.

\subsection{Visual Encoding Recommendation}

Historically, the visualization community has frequently used the term \textit{recommendation} in the context of recommending one or a set of alternative visual encodings for a specific combination of data types.
Tableau ShowMe~\cite{mackinlay2007showMe}, Voyager~\cite{wongsuphasawat2017voyager}, Draco~\cite{moritz2018draco}, Data2Vis~\cite{dibia2019data2vis}, and VizML~\cite{hu2019vizml} are examples of such recommendation engines. 
Tremendous progress has been made to handle increasingly complex combinations of data types, to incorporate perceptual effectiveness scores, and to leverage machine learning models~\cite{saket2018beyondHeuristics}. However, much remains to be done until automated visualization design reaches the sophistication of manually crafted visualizations and dashboards, that include dynamic data streams, data transformations, multi-view interactions, and custom annotations which collectively involve countless human design decisions~\cite{sarikaya2018dashboards}. In contrast, the long-term goal of our effort is to develop a traditional recommender system to surface existing viz workbooks that have been created by users and are stored in VizRepos, and not the recommendation of visual encodings for given datasets.

\subsection{Semantic Text Similarity Models and Evaluations}

Semantic text similarity tasks are common in the field of natural language processing to improve algorithms for applications such as machine translation, summarization, or question answering~\cite{agirre2016semeval,o2008comparativeStudySimMeasure}. However, these tasks are often based on curated datasets, such as sentence pairs, that do not reflect the special characteristics of the text that is embedded in viz workbooks and results might not be generally applicable.

Previous work compared human judgments with model predictions for the task of semantic text similarity. Towne et al.~\cite{towne2016measuringSimilaritySimilarly} compared human judgements with LDA predictions based on multi-paragraph documents. Colucci et al.~\cite{colucci2016evalItemItemSimilarity} compared human judgements with TF-IDF in the context of movie recommendations. In contrast, we compare a broader range of models including word embeddings.

\subsection{Evaluation of Recommendation Algorithms}

Many approaches have been proposed to evaluate recommendation algorithms, both in general and for CBR systems specifically, which in turn dictate the model selection and refinement process~\cite{shani2011evaluatingRecSys}.

A/B testing allows us to compare algorithms in the real world and is usually the most desired method but is costly and often not feasible, particularly if there are multiple algorithms to test.
Instead, offline metrics are frequently used to evaluate algorithms~\cite{levy2013offlineEval}. Utility metrics, such as coverage, scalability, or computational performance, can be evaluated independently while accuracy metrics require ground truth data that is not always available. Moreover, several studies~\cite{mcnee2006accurateNotEnough,knijnenburg2012explainingUxRecSys,konstan2012fromAlgToUser} demonstrated that these standard metrics that are used to evaluate recommender systems are often not adequate to reflect user expectations and a more user-centric evaluation is necessary.

User studies either assess the users' satisfaction with recommendations in the context of a specific interface directly~\cite{du2018sequenceRecSys} or exclusively evaluate the underlying algorithm. Our work focuses on the latter and specifically about the similarity model as the main building block towards content-based visualization recommendations.

Several studies investigated how human similarity perception aligns with the algorithmic notion of similarity for specific domains~\cite{lee2010crowdsourcingMusicSimilarity,li2016faceSimilarityJudgements,tirilly2012imageSimilarity}. Winecoff et al.~\cite{winecoff2019psychSimilarItem} collected relative similarity judgements to compare the Jaccard similarity index with a new metric that accounts for psychological factors. In a similar spirit, we also conducted a crowdsourced two-alternative  forced choice study but our task focused on the semantic similarity between semi-structured text snippets instead of the visual appearance of fashion items.

More closely related to our approach is the work by Yao \& Harper~\cite{yao2018judgingSimilarity} who evaluated several CFR and CBR methods for movie recommendations with similarity ratings by users. The main finding was that CBR algorithms outperform CFR methods. The authors asked participants to rate the similarity between pairs of familiar movies on an absolute scale while we chose to collect relative similarity judgements that are more robust and reliable~\cite{demiralp2014perceptualKernels,li2016faceSimilarityJudgements}. In addition, they showed the title and movie poster of familiar movies while our study interface shows all the text to participants that is also used for model predictions.


\section{VizCommender Proof-Of-Concept}\label{sec:proof-of-concept}

We now describe our proof-of-concept pipeline, VizCommender, that we created to investigate different similarity measures and to get stakeholder feedback about visualization recommendations. 

We use a VizRepo with 4,698 Tableau Public workbooks as input data and extract text elements from the viz specifications that are informative about the underlying topic. We convert these bag-of-words into numeric vector representations and compute the similarity between them using NLP techniques in order to generate workbook recommendations. The final version of the underlying model is informed by the results of our crowdsourced user study: we decided to instantiate a similarity model based on LDA and Jensen-Shannon divergence, as described in~\autoref{sec:prototype-instantiation}.

We preprocess these similarities and when users browse in the prototype front-end, content-based recommendations are displayed, as described in the following section.

\subsection{User Interface}\label{sec:user-interface}

Initially, the VizCommender interface displays an overview of many workbooks arranged in a scrollable grid view, as shown in~\autoref{fig:explorer-overview}. Each \textit{workbook preview} consists of a thumbnail, title, author name, and date. If a workbook contains a dashboard, the screenshot of a dashboard is shown as a thumbnail to make it immediately apparent. An auto-suggest search box enables users to find workbooks by keywords or author names. We use TF-IDF to extract keywords from viz specifications and display the most important ones across all loaded workbooks as tags at the top of the page. By clicking on a tag, users can further drill-down.

Clicking a workbook preview opens the \textit{quick view sidebar} with further details about that workbook, including custom recommendations.


Users can open the interactive detail view from the sidebar or through a double click on the workbook preview. The detail view, shown in~\autoref{fig:explorer-detail}, contains a panel at the bottom of the screen that can be expanded to reveal workbook recommendations similar to the quick view. 
A tab navigation allows users to explore recommendations based on three \textit{similarity facets}:
(1) related workbooks, (2) similar versions, and (3) workbooks with similar data. We provide further details about these similarity facets in~\autoref{sec:prototype-instantiation}. The goal is to provide groups where users can choose from instead of creating a single all-purpose list.
Additional recommendations can be loaded by horizontally scrolling within the panel.

\subsection{Example Recommendation \& Usage}\label{sec:vizcommender-interface}

An example workbook about Olympic medals is selected in~\autoref{fig:explorer-detail}a. The recommendation panel in ~\autoref{fig:explorer-detail}b shows two related workbooks that are sufficiently different but similar in topic. The first four recommendations from the \textit{similar data} facet are shown in ~\autoref{fig:explorer-detail}d. One workbook is included in both facets which illustrates that the models lead to related results although one is trained on all textual content and the other one is only trained on data column names.

We used VizCommender regularly ourselves and with our collaborators to better understand and debug model results. For example, we implemented recommendations for individual sheets and workbooks and only after several chauffeured demo sessions we concluded that workbooks are the primary object of interest.

VizCommender also helped us to tune threshold values for the similarity facets (see~\autoref{sec:prototype-instantiation}). We identified imprecise recommendations that have no topic-relationship to the reference which can not be avoided because of the nature of text data and the relatively small VizRepos but mitigated through higher thresholds.


\section{Data Extraction and Feature Engineering}\label{sec:data-characterization}

We now describe the characteristics of VizRepos and the input features that we extract and use in our approach, and describe the intrinsic challenges of the data in a CBR context.



\subsection{Visualization Text Extraction}

A viz workbook specification contains a number of text objects: workbook name, sheet names, titles, axes labels, captions, annotations, and data source column names. All of these objects can contain human readable text that may, but does not always, capture meaningful semantic information about the topic of the visualization. We use the collection of words and phrases to construct a \textit{document} in order to apply NLP techniques to compute similarities and classifications between documents.
After preprocessing, we store the data in a PostgreSQL database to ease querying and further analyses.

For each workbook, we extract and preprocess the text from the XML specification. We remove numbers, punctuation, strings with less than three characters. We apply a custom stop-word filter to eliminate text such as "Number of Records" that is auto-generated by Tableau for all data sources. We created a short list of 354 generic stop-words by analyzing frequently but non-informative words or phrases that occur in Tableau VizRepos (see supplemental material). More extensive corpus-specific filters are beyond the scope of this work; previous work~\cite{schofield2017pullingOutStops} indicates that stop-word removal might offer only superficial improvement. We also annotate the text with part-of-speech tagging and lemmatize it using \textit{NLTK WordNet}~\cite{bird2009nltk}.

\subsection{Leaving Out Visual Encodings}\label{sec:leave-out-visual}
In this work, we chose to focus on basing recommendations on the subject matter of a visualization or workbook and not the visual style. Thereby, we ignore all visual encoding specifications such as mark types, colors, reference lines, and layout properties. This decision was informed by in-depth discussions with our collaborators and the prioritization of information seeking. This task is highly focused on finding informative workbooks around a user's specific analytical questions. For example, while looking at a 2019 report, a CBR system might also recommend the related 2018 report for comparison and it is not relevant whether the workbook contains a scatter plot or a bar chart. We also verified this expert guidance informally through our prototype. An early version used an adjustable weighted combination of visual and topic features, but we found that directly using visual encoding features added noise to the model when the task is information seeking in VizRepos. We saw that some workbooks are visually perceived similar although they use different data and address distinct topics, and conversely identical information may be visualized fundamentally differently. Although visual styles are not directly incorporated, we do extract \textit{chart-visible} text elements, such as axis titles or annotations, that may reveal information about the underlying topic.

A few very large and public-facing VizRepos, such as Tableau Public or Power BI's Data Stories Gallery, are unusual cases that are focused on entirely different tasks than is typical in an enterprise setting. One such task is searching for design inspiration, which would indeed require making visual encoding specifications first class citizens. However, that task is beyond the scope of this paper.

\subsection{Leaving Out Underlying Data}
All viz specifications inherently also include references to the underlying data being visualized, and this data includes additional semantic information that could be incorporated. However, early in the project, we made the fundamental decision to not access a workbook's underlying raw data. In enterprise settings, viz workbooks are often not just built upon simple CSV files and instead are powered by large scale data warehouses or live data streams. Including all such data content for recommendation purposes would quickly lead to computational scalability challenges. 
Given this concern, we decided the computational cost required to process raw data for each workbook to be prohibitive for large scale VizRepos, and for enhanced scalability chose to focus on the more accessible data within the viz specification itself. Incorporating samples of the raw data offers an interesting avenue for future work that can be informed by our results.

\subsection{Data Challenges}\label{sec:data-challenges}

\begin{figure}
 \centering
 \includegraphics[width=0.98\columnwidth]{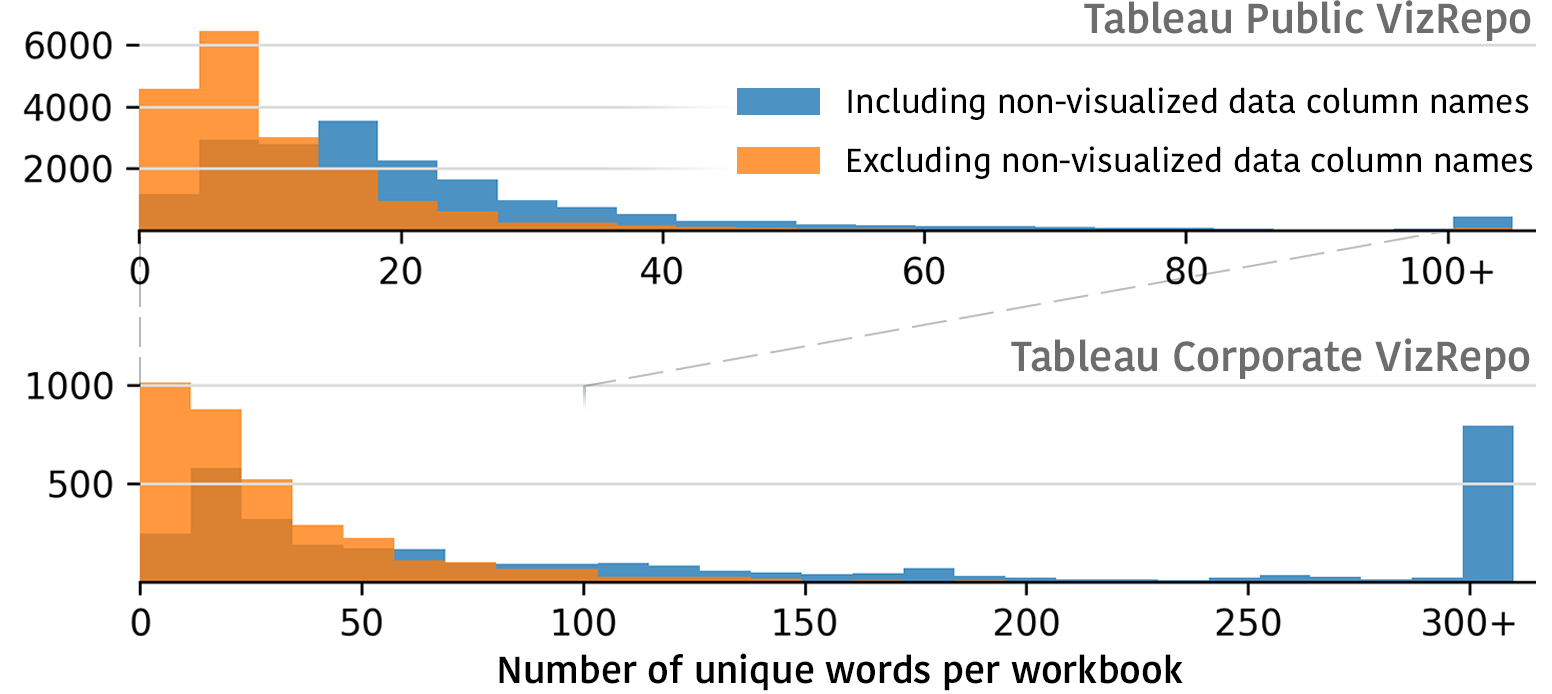}
 \vspace{-8pt}
 \caption{Unique words per workbook with and without data column names. Incorporating column names significantly increases the document size, most notable for the corporate VizRepo, and can be a strong signal when comparing the topic similarity between workbooks.}
 \label{fig:unique-word-count}
 \vspace{-12pt}
\end{figure}

Based on the two previously mentioned sample VizRepos, we identified five challenges in leveraging text content from viz workbooks:

\vskip 5pt
\noindent \textit{C1: Workbooks and Nested Visualizations.}
For many business intelligence applications such as Tableau, the dashboard is the primary visual artifact being created. A dashboard is typically made of multiple related visualizations, that may share more than one data source.
Additionally, it is common for any single visualization to contain faceted nested visualizations in order to create trellis views or small multiples. The result is a potentially complex hierarchy of visualization specifications. A given workbook can contain an arbitrary number of such complex dashboards, compounding the structural complexity.

\vskip 5pt
\noindent \textit{C2: Very Limited Text.}
The text found in most sheets generally consists of fragments of a few words, such as titles or labels. Less frequently, a visualization author might include captions or annotations of lengthier text, but not usually something approaching the length of an actual text document. See~\autoref{fig:unique-word-count} for a distribution of unique words per workbook. This characteristic is somewhat similar to the problem of classifying tweets~\cite{sriram2010shortTextTwitter}, but is even more extreme: the title and label text may or may not be strictly related, and certainly do not have the same natural language relationships that a typical grammatical sentence would have. An additional problem with titles and labels is that Tableau provides auto-generated text by default which simply re-iterates the field names being visualized and provides no additional semantic information.

\vskip 5pt
\noindent \textit{C3: Incomplete Workbooks.}
In both VizRepos, we found a substantial number of low quality workbooks that were incomplete, work-in-progress drafts, or even completely empty.

\vskip 5pt
\noindent \textit{C4: Multiple Versions.}
In both repositories we found numerous examples of exact-duplicate or near-identical workbooks. This challenge was particularly striking on Tableau Public which is often the system used for teaching Tableau to students and data scientists. The result is that there are many near-duplicate workbooks related to a particular course curriculum, or on-line design challenge as for example \mbox{MakeoverMonday}~\cite{kriebel2018makeoverMonday}. A similar problem exists with periodic reports such as corporate finance reports that may occur quarterly, or annually. Choosing the most relevant from these sets of nearly identical workbooks is a major challenge and strongly linked to the individual context.
Someone working on a 2020 report might want to see a reference to the 2019 report, but someone working on a class project may not want to see many other very similar examples of the same use case.


\vskip 5pt
\noindent \textit{C5: Out-of-Vocabulary Words.}
Most visualizations have a domain-specific nature reflected in their data and text. This domain specificity can make the use of pre-trained NLP models less effective since the text is often expressed in proprietary company-internal vocabulary or a very domain-specific language. Although VizRepos frequently contain thousands of workbooks, the corpus size is too small to train custom word embeddings that accurately capture word representations~\cite{mikolov2013distributedRepresentations}. Further, column names are often abbreviated or totally cryptic to any user other than the database administrator, and such naming enigmas carry over to default labels generated from such field names.

\subsection{Feature Selection and Data Filtering}

In Tableau repositories, a workbook is the primary artifact that combines visualizations targeting an analytical question or to communicate data around a specific subject. Broadly comparable to Excel files, all sheets within a workbook typically focus on the same topic and individual views are building blocks for dashboards. In this work we propose only workbook to workbook comparisons, for an achievable scope that eliminates the need to decompose nested visualizations and cross-references to individual views (C1). We found only a small number of workbooks that contain a mixture of distinct topics and speculate that these cases primarily occur when users learn Tableau and experiment with features and different data sources.

Analyzing an entire workbook allows more text to be extracted than from a single sheet and thus helps alleviate some of the issues that arise from feature extraction of a single visualization (C2, C5). We also use the column names from data sources as additional text data. \autoref{fig:unique-word-count} shows how the number of unique words per workbook increases when column names are incorporated. This choice includes data that may or may not be used in any visualizations, so that the semantic content of the data source itself can help inform us about the topic it describes.
While this approach can frequently result in repeated text fragments, it is analogous to human-authored text documents with repeated words and phrases. Standard methods for normalizing word frequency can be applied to rigorously and meaningfully account for such repetitions.


For recommendation purposes, we exclude workbooks with no visual marks (C3) but we include mostly-complete workbooks that contain a few partial sheets because those could be remnants from a data analysis process that have not been explicitly removed. While it would be highly desirable to filter out misleading and deceptive visualizations from recommendations, automatically assessing the quality of visualizations remains one of the major challenges in our research field~\cite{kindlmann2014algebraicVis,mcnutt2020VisMirages} and is not the focus of this work.

Many of the NLP models we use generate poor or unreliable comparisons for very small number of features (that is, a too-small \textit{bag of words}) in one or both workbooks being compared. Hence, after feature extraction and stop-word filtering, any workbook with less than 10 relevant words is also removed from consideration (C3).


\section{Viz-to-Viz Similarity}\label{sec:viz-to-viz-similarity}

We propose to use textual similarity for creating visualization recommendations and briefly summarize four NLP models that we considered.

\subsection{Pairwise Comparisons}

In order to create a ranked list of recommended visualizations or workbooks, we need some mechanism to make comparative judgements. 
In this work, we focus on the use case where a user looks at a reference workbook and wants to discover other workbooks that lead to additional insights. The basis for this computational model can be distilled to comparisons between pairs of items. As we will discuss in~\autoref{sec:advanced-recommendations}, more advanced models can build upon this similarity measure and incorporate multiple workbooks and user profiles.

We chose to pursue an approach that uses textual features extracted from pairs of visualization workbook specifications and computes the similarity between the two sets of features. In line with other CBR systems, our assumption is that similarity is a surrogate for relevance and a crucial building block for a recommender system, even though for a full-fledged instantiation many other factors such as diversity, serendipity, novelty, and trust must also be taken into account~\cite{ge2010beyondAccuracy}.


\subsection{NLP Models}\label{sec:nlp-models}

We now review a set of methods to compute pairwise similarity scores based on textual features. A broad range of NLP techniques exist that have been widely adopted for many different applications in a variety of domains that are also applicable to CBR contexts. 

We conducted a literature review to assess the suitability based on the special combination of data characteristics in VizRepos and narrowed down the selection to four general models, each composed of a document representation method and a distance measure. The models we use in our experiments are TF-IDF, LSI, LDA, and document embeddings Doc2Vec and GloVe, summarized below. We chose TF-IDF as a simple but powerful baseline model. Our collaborators were particularly interested in LSI as they were considering using it in a hybrid recommender system. LDA is popular for topic modeling but 
also commonly used to assess text similarity. Recent document embedding approaches surpass these traditional methods for many NLP tasks, so we also tested them. 

In addition to these widely used models, which we chose as an appropriate first step, many potentially useful alternatives exist that could be explored as future work, particularly word/document embedding techniques~\cite{peters2018elmo,devlin2018bert,bojanowski2016enriching} that have received significant attention recently. ELMo~\cite{peters2018elmo} and BERT~\cite{devlin2018bert}, for example, can better capture out-of-vocabulary words because embeddings are learned for n-grams but their primary strengths of incorporating the surrounding context of words is more applicable to actual text documents with sentence structures than for viz specifications.

We use cosine similarity, one of the most commonly used distance measures in vector space, for all models except LDA. Following previous work~\cite{beck2018readNext}, we decided to use Jensen-Shannon divergence~\cite{lin1991divergence} for LDA because it is more suitable for probability distributions.

All NLP models were implemented in Python based on the powerful and popular \textit{gensim} software package~\cite{rehurek2010gensim}.

To determine which of those selected models is a good fit for our problem we conducted an informal investigation of all models followed by a crowdsourced user study to systematically assess the alignment between model predictions and human judgements (see~\autoref{sec:study}).

\vskip 4pt
\noindent\textbf{\textit{Model:} Term Frequency–Inverse Document Frequency (TF-IDF)}\\
\noindent\textbf{\textit{Metric:} Cosine Similarity} \\
\noindent TF-IDF~\cite{robertson2004tfidf} is an extension of a simple bag-of-words approach that considers a term’s frequency (TF) and its inverse document frequency (IDF) to promote terms that are unique across the corpus and to weed out common language. In essence, documents are vectorized by the relative importance of each word and TF-IDF functions as a more advanced stop-word filter. After computing a numerical TF-IDF-weighted score for each term, the result is a sparse two-dimensional document-term matrix. We use cosine similarity to compute the similarity between vectors in the high dimensional space; other suitable similarity measures~\cite{agarwal2017similarityMeasures} could also be used.
TF-IDF is a simple and efficient algorithm but does not capture word semantics and online learning of new documents requires revisiting all the previous documents.

\vskip 4pt
\noindent\textbf{\textit{Model:} Latent Semantic Indexing (LSI)}\\
\noindent\textbf{\textit{Metric:} Cosine Similarity} \\
LSI~\cite{deerwester1990lsi}, also referred to as \textit{latent semantic analysis} (LSA), uses truncated singular value decomposition to analyze relationships between terms. The assumption is that words that appear together in a document have a similar meaning.
An LSI model, trained on a TF-IDF weighted term-document matrix, produces low dimensional document representations that can be compared using cosine similarity. The number of latent dimensions is the most important hyper-parameter and can be difficult to determine. We test 15, 30, 75, and 150 dimensions.

\vskip 4pt
\noindent\textbf{\textit{Model:} Latent Dirichlet Allocation (LDA)}\\
\noindent\textbf{\textit{Metric:} Jensen-Shannon Divergence (JSD)} \\
The idea behind LDA~\cite{blei2003lda} is that each document contains a mixture of latent topics that are found throughout the corpus and each topic is composed of a set of words. LDA is a probabilistic model that is first trained on bag-of-words of documents and then computes a probability distribution over \textit{k} topics for each document. We use the inverse of JSD to compare the similarity between topic distributions. LDA is the most popular technique for topic modeling but also frequently used for semantic similarity tasks.
As with LSI, the number of topics must be predefined and an optimal configuration is difficult to determine. We probe the model with 15, 30, 75, and 150 topics.

\vskip 4pt
\noindent \textbf{\textit{Model:} Document Embeddings}\\
\noindent \textbf{\textit{Metric:} Cosine Similarity} \\
We distinguish between three variations, \textbf{Doc2Vec}, GloVe Pre-Trained (\textbf{GloVe-Pre}), and GloVe Transfer Learning (\textbf{GloVe-TF}).

Doc2Vec~\cite{le2014doc2vec} generalizes the Word2Vec word embedding method to documents. Word2Vec is a neural network that produces numeric representations for each word by incorporating the word context in order to capture extensive semantic relationships, such as synonyms, antonyms, or analogies. Doc2Vec extends this model to sentences or whole documents to create fixed-length document representations. The main challenge is to train Doc2Vec on a sufficiently large corpus, as is the case with most machine learning methods. We use the \textit{Distributed Bag Of Words} (DBOW) approach on a large set of viz specifications and test the model with 100 and 300 dimensional feature vectors.

GloVe~\cite{pennington2014glove} is an alternative to Word2Vec to train word embeddings that uses a co-occurrence matrix of the corpus. Several pre-trained models have been made available. For GloVe-Pre, we use Stanford's model~\cite{pennington2014glove} trained on 400K vocabularies from Wikipedia and Gigaword (vectors with 100 and 300 dimensions). In order to get one vector per document, we compute the average word embedding. This model could be further tuned by using more advanced techniques to combine individual word embeddings such as Word Mover's Embedding~\cite{wu2018wordMoverEmbedding}.

While many pretrained models based on news or Wikipedia articles have been released, they are often not adequate for domain-specific documents and lack important vocabulary. For GloVe-TF, we use the pre-trained GloVe-Pre model and further train it on the viz specifications from our corpus, similar to Doc2Vec.

In case of Doc2Vec, we generate five permutations with different word orderings to improve the performance. The strength of these embedding models to learn semantics through surrounding words is significantly impacted because the extracted text from viz specifications resembles an unordered list of tags instead of actual text documents (C2). See~\autoref{fig:unique-word-count} for a distribution of words per workbook.

We also used our prototype, VizCommender, to analyze the model output in a CBR context directly. This subjective inspection allowed us to see if the recommendations make sense for domain experts.


\section{Crowdsourced Similarity Judgements}\label{sec:study}

We now describe the user study in which we collected human judgements from a semantic text similarity task, in order to use the resulting data to inform the selection of appropriate NLP models.

\begin{figure}
 \centering
 \includegraphics[width=\columnwidth]{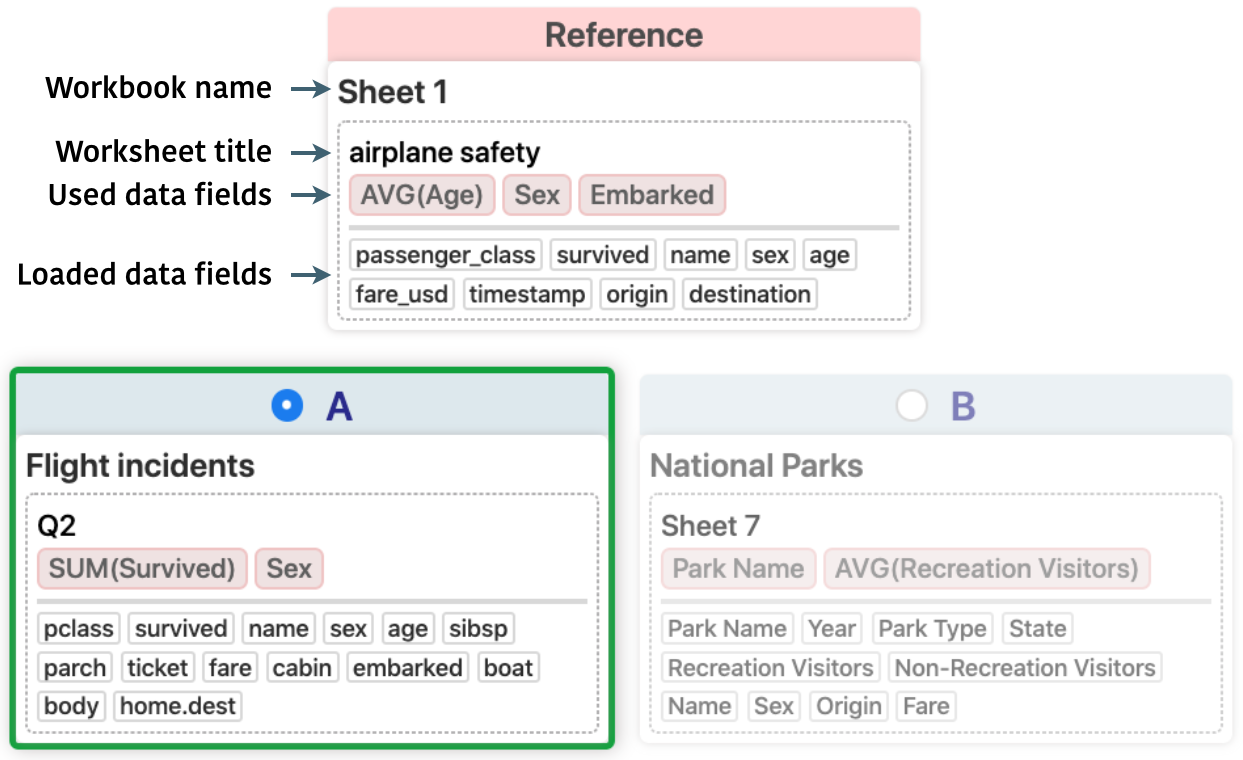}
 \caption{Annotated user interface for the semantic text similarity task. Participants select one of two alternatives that is most similar to a reference.}
 \label{fig:vis-triplet}
 \vspace{-12pt}
\end{figure}

\subsection{Method}

To assess the alignment between human judgements and our algorithmic notion of similarity, we conducted a crowdsourced two-alternative force choice (2AFC) experiment using the Mechanical Turk (MTurk) platform.

\vskip 5pt
\noindent \textbf{Human similarity judgements:} Participants were presented sets of three documents (triplets) and asked to select one of two alternatives that is most similar to a reference document, as shown in~\autoref{fig:vis-triplet}.
Several studies~\cite{demiralp2014perceptualKernels,li2016faceSimilarityJudgements} have shown that triplet matching and relative similarity judgements are more reliable and robust across participants than pairwise judgements where subjects are asked to rate the perceived similarity on a numerical scale. An early pilot study using pairwise ratings confirmed these limitations by showing that participants interpret the absolute scale differently, an ambiguity that does not exist in relative similarity judgements.

\vskip 5pt
\noindent \textbf{Experimental stimulus:}
As described earlier, our objective is semantic text similarity and not the visual style of visualizations. Therefore, participants were only shown text extracted from viz specifications with no indication of the layout or chart type (see~\autoref{fig:vis-triplet}). 
To make judgements less challenging and allow crowdsourced participation, we reduced the scale of each document by including only content from one visualization (sheet) and the workbook title. Using only these features yields identical results to single-sheet workbooks, which constitute roughly one third of all workbooks across the corpora that we analyzed.

\vskip 5pt
\noindent \textbf{Model similarity judgements:}
Analogous to collecting human judgements, we ran the same experiment with NLP models (TF-IDF, LSI, LDA, Doc2Vec, and GloVe) to determine if the algorithmic notion of similarity mimics human judgement and if there are significant differences between models. To imitate the judgement task, models produce predictions by independently computing the pairwise similarity scores between the reference document and one of the alternatives. The model compares the two scores and the alternative with the higher score is deemed to be more similar.

\subsection{Data}

We sampled data in a multi-stage process from the corpus of 18,820 Tableau Public workbooks. 

\vskip 5pt
\noindent \textbf{Visualization selection}: We randomly selected 4,698 workbooks that were viewed at least twice on public.tableau.com and recognized to use English language. We extracted textual features from 19,575 visualizations within them and filtered out visualizations with no visual marks. We also discarded sheets that contain less than 10 or more than 200 words, including the workbook title, which resulted in a total number of 15,482 visualizations.

\vskip 5pt
\noindent \textbf{Triplet pool generation}: We sampled triplets based on TF-IDF and LDA. We used TF-IDF as a baseline model to calculate pairwise similarity scores (0-1) between the 15,482 visualizations before we randomly assembled the document triplets, under specific conditions: First, we considered only scores between 0.15 and 0.9 to exclude highly dissimilar and nearly identical visualizations. Second, we set the delta between the two reference-alternative scores within a triplet to a minimum of 0.45 to ensure that alternatives are sufficiently different on a syntactic level. Asking participants to rate identical or almost identical alternatives would be less insightful. Third, to avoid a scenario of too many examples revolving around the same topic, such as customer sales or basketball statistics, we used stratified sampling based on LDA topics and added each sheet to at most two triplets. All thresholds were informed through pilot testing and by gradually examining combinations of similarity scores. This stage led to a pool of 4,211 triplets.

\vskip 5pt
\noindent \textbf{Triplet selection}: Based on the available triplets, we randomly sampled 135 candidates and manually inspected each for mixed-language content and extensive domain-specific vocabulary, that could not be filtered automatically. A new triplet was sampled for each rejection. In total, we examined 169 triplets to get the 135 retained.

\vskip 5pt
\noindent \textbf{Model training data}: All models are trained on 96,439 viz sheet specifications from the Tableau Public VizRepo that are not included in the 4,211 sampled triplets.

\subsection{Procedure}

We recruited participants from Amazon's MTurk. Using an online crowd of workers to complete human intelligent tasks (HIT) is a common practice in visualization~\cite{heer2010crowdsourcing}, and has been used in the past to collect similarity judgements~\cite{winecoff2019psychSimilarItem,lee2010crowdsourcingMusicSimilarity}. 

Eligible workers with an approval rate greater than 98\% and more than 5000 approved HITs were redirected to our custom web page. The page provided the consent form, a screening task, and detailed instructions, specifically pointing out that decisions should be based on semantics and the gist of a document instead of individual word overlaps. The emphasis is on correctness instead of speed. The screening test involved the rating of a triplet where the semantic relatedness between the reference and one of the alternatives is immediately obvious.  

We divided the 135 triplets into three equal batches of 45 triplets that were each rated by 25 participants. The triplets were shown in random order and the position of the alternatives was randomized. The second and tenth triplet were repeated at the end again, to assess the intra-rater agreement, which led to 47 required judgements. After finishing all ratings and the post-questionnaire, participants were provided a unique code to enter on MTurk. We compensated workers with a flat rate honorarium of US\$ 4.00 on completion of the study. This amount was defined based on internal pre-tests indicating that a session would require roughly 20-30 minutes. A participant who completed the task or was deemed ineligible was prevented from attempting the task a second time.

\subsection{Limitations}

The goal of this study was to better understand the appropriateness of various NLP models for similarity comparisons. Participants were only shown text extracted from viz specifications that is also used for model predictions. Showing visualizations instead may have led to different results but our focus is on the subject matter and therefore we excluded the visual encodings to reduce factors influencing the judgements.

We see this semantic text similarity study as a necessary step towards creating a content-based recommender system and not as a replacement of a summative evaluation of recommendations.  Many other factors play an important role when evaluating recommendations in an application, such as user interface design, trust, and familiarity, which would be interesting to examine but these aspects were not the immediate objective of this work. 


\section{Results}

\begin{figure}
 \centering
 \includegraphics[width=\linewidth]{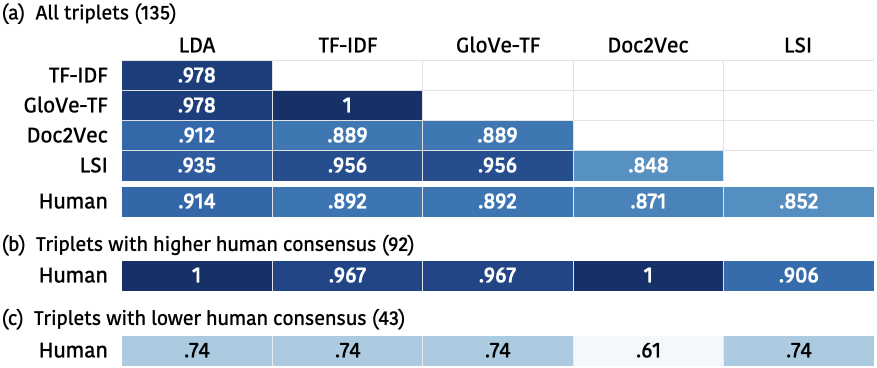}
 \caption{Agreement between model predictions and human judgements for the semantic text similarity task. Each cell shows the Fleiss' Kappa agreement between two models or between the human majority vote and the model prediction.}
 \vspace{-4pt}
 \label{fig:sts-agreement-scores}
 \vspace{-12pt}
\end{figure}

We describe the results of our experiment and discuss how they inform the selection of an appropriate model for the similarity measure.

We implemented an interactive visual analysis tool to both detect suspicious crowdsourced data and to better understand the results and the implications for the model selection, similar in spirit although different in details to Kachkaev et al.~\cite{kachkaev2014glyphsExploreCrowd}. Screenshots of this tool, triplet input data, and human judgements are in supplemental material.






\subsection{Human Judgement}

A total of 89 participants accepted the MTurk HIT. Of those, 75 (n=75) were deemed eligible and completed the experiment; 5 workers responded incorrectly to the screening task at the beginning and 8 workers did not complete all required steps. Before the release of the study, we divided the set of 135 triplets randomly into three batches of 45 triplets each. Once 25 workers completed a batch, we took it offline. We also removed one participant from the analysis who responded implausibly fast and thus reran the batch with a new worker.

One of the researchers in our team labeled all 135 triplets beforehand to create a gold standard to be used to filter out obviously poorly performing participants. Our analysis tool shows the responses of each participant displayed along a timeline and color-coded based on the agreement with the gold standard. We also tracked if participants primarily pick alternatives on the left or right side, indicating a lack of attention to the intended task, but no participant was excluded from the analysis based on these two mechanisms.

Of the 75 participants, 57.3\% were male, 41.3\% were female, and 1.3\% non-binary. Further details about the demographics, such as age distribution and educational background, are included in Supp. Sect. 5.

We use Fleiss' Kappa $\kappa$~\cite{fleiss1971fleissKappa} that quantifies the agreement between a fixed number of two or more raters in order to compare model predictions with human judgements. Fleiss' Kappa can range from -1 (no agreement) to +1 (perfect agreement) and is comparable to Cohen's inter-rater agreement~\cite{cohen1960cohenKappa} for the case of two raters.
The inter-rater agreement between all participants (mean Kappa values of all three batches) is $\kappa=.544$, a moderate agreement according to the commonly used interpretation of Landis and Koch~\cite{landis1977kappaInterpretation}. 

We duplicated 2 triplets to assess the intra-rater agreement. 56 participants (74.7\%) have a perfect agreement, 17 (22.7\%) rated 1 triplet differently and 2 reversed their judgements for both repetitions. The relatively large group of people who changed their opinion could be an indicator that, for these triplets, both alternatives are related to the reference according to human judgement despite the clear preference in the baseline TF-IDF model. 
37 triplets (27.4\%) were judged unanimously by all participants. One triplet has an equal number of votes for both alternatives. There is also a strong agreement between the participants and the gold standard, although the majority of participants disagreed in 4 cases.

In light of this analysis, we classify triplets into two groups based on human consensus: 92 \textit{high consensus} triplets with at least 80\% agreement between all participants, and 43 \textit{low consensus} triplets with less than 80\% agreement or disagreement between the gold standard and the majority vote. A detailed examination of low consensus triplets confirmed that these involved non-trivial alternative choices.

\subsection{Model Comparison}

We now compare the models with the human consensus using the majority rule (alternative chosen by the majority of
the participants), with results summarized in ~\autoref{fig:sts-agreement-scores}. Detailed results for varying model parameters are included in Supp. Sect. 5.4.

The LDA model aligns best with human judgements for all triplets, regardless of low or high human consensus. \autoref{fig:sts-agreement-scores} shows agreement scores for the LDA model with 150 topics but lower number of topics (30 and 75) achieved comparable results.

GloVe-Pre, the pre-trained word embedding model, and GloVe-TF, the extension that is further trained on extracted text from viz specifications, made identical predictions. We conjecture that the relatively small number of 96,439 viz specifications used for online learning is insufficient for GloVe-TF to achieve an additional gain.

LSI performed worst, regardless of the number of dimensions. 

Model predictions align almost perfectly with judgements for triplets with higher human consensus (\autoref{fig:sts-agreement-scores}b). In contrast, predictions for triplets with lower human consensus diverged the most from human judgements, most notably for the newly trained Doc2Vec model (\autoref{fig:sts-agreement-scores}c). These results are an indication that the Doc2Vec embeddings do not capture the underlying semantics sufficiently and more training data would be necessary to do so.

In general, the results demonstrate that all models are nearly on par except for Doc2Vec and LSI. The result that simpler models, such as LDA or even TF-IDF, are not substantially worse than more complex models confirms that previous findings~\cite{chen2016compareTfidfLdaParVector} apply to this task setting. 

Although each model can be further tuned and more sophisticated NLP techniques can be applied, the results indicate that off-the-shelf models can be leveraged as a baseline despite the challenging data characteristics discussed in~\autoref{sec:data-challenges}. Although our  comparison with human judgement is a reasonable first step, it is only one signal and other factors, such as performance or explainability of results, should also be considered as next steps when choosing an appropriate model.

\subsection{Prototype Instantiation \& Similarity Facets}\label{sec:prototype-instantiation}

We decided to instantiate a similarity model based on LDA and Jensen-Shannon divergence in our final proof-of-concept system. Although other models also had high performance, including document embeddings based on GloVe, the characteristics of LDA that led us to choose it are its relative simplicity and its utility for other user interface use cases such as showing topic membership as well. Although other approaches are even more appropriate for online learning, it is feasible with LDA~\cite{hoffman2010onlineLearningLDA} and medium-sized VizRepos can be re-processed regularly. We discuss further scalability implications in~\autoref{sec:limitations}.

Our requirements elicitation process revealed that a single numeric similarity score is not adequate for the diverse tasks that should be facilitated by a recommender system. We have identified three types of recommendations that we refer to as \textit{similarity facets} where a text-based similarity model can be applied. All three facets use the same underlying model but require a slightly different optimization of feature selection and score thresholds. We tuned these thresholds manually by reviewing example recommendations in the interface (see~\autoref{sec:vizcommender-interface}).

\vskip 3pt
\noindent
\textbf{F1. Related Workbooks}: This facet uses all extracted text objects from workbook specifications including column names and computes a pairwise similarity score between 0-1 with the LDA-based model. The objective is to find workbooks that are semantically related but not near-duplicates (C4). We ultimately chose 0.65-0.9 on a scale from 0-1. Tuning these cut-off thresholds leads to a trade-off. By expanding the range of acceptable scores, we increase the diversity of results and the chance for serendipitous discoveries while we simultaneously increase the risk of including less relevant workbooks.

\vskip 3pt
\noindent
\textbf{F2. Similar Versions}: In contrast to F1, users are sometimes specifically interested in seeing different versions of a workbook (C4), a task that can be expedited by a recommender system identifying exactly the near-duplicates. This facet uses the same model and similarity scores as F1 but uses a different cut-off threshold that is set very high ($\geq$ 0.9).

\vskip 3pt
\noindent
\textbf{F3. Similar Data}: In some scenarios, many workbooks are semantically similar or cover the same topic. To support more targeted recommendations, we propose a facet that only takes columns names as input and ignores other textual features in order to reveal workbooks based on related data sources. Although the underlying data might not be identical, we hypothesize that a certain degree of overlap between column names is an indicator of semantic data similarity. We also use LDA for this facet but train a separate model with different input data, and choose a high threshold of $\geq$ 0.9.

\vskip 5pt
\noindent 
We surface these similarity facets through our \textit{interactive recommender system}~\cite{he2016interactiveRecSys} prototype that allows users to interactively specify their intent while browsing, as described in~\autoref{sec:user-interface} and illustrated in~\autoref{fig:explorer-detail}.

\section{Discussion}

In this section we discuss limitations, other use cases, and future work.

\subsection{Limitations of Tableau Repositories}\label{sec:limitations}

Our investigations were based on two Tableau repositories, one public and one corporate. In our crowdsourced experiment with external participants and the VizCommender interface examples shown in this paper, we rely on the public one containing a sample of workbooks obtained from Tableau Public, because these workbooks are publicly available and not restricted by non-disclosure agreements.
The size of this sampled VizRepo corresponds to typical corporate repositories with a few thousand workbooks, although we are aware of much smaller and larger VizRepos in regular use.

The challenge with a small number of workbooks becomes one of training meaningful topics for the LDA-based similarity model. The results of our study demonstrate that a GloVe model that is pre-trained on external data represents a viable alternative.

The challenge of very large VizRepos include scalability problems, although we note that the similarities between viz specifications can be pre-computed offline. One way to address this issue would be to use a two-step process. First, the MinHash~\cite{broder1997resemblanceContainmentDocs,chum2007scalableNearIdenticalImageDetection} algorithm can be applied to compare a large number of viz specifications quickly. Second, a more precise NLP model, such as LDA and Jensen-Shannon divergence, can be applied to a small set of items that is selected in the first step.

In addition to the raw size of a VizRepo, for our proposed viz-to-viz similarity model to function there must be a large enough number of words within its workbooks to discriminate between them or else the recommendations become imprecise. In the VizCommender, we preclude workbooks with less than 10 words from being considered for recommendations. Our suggested text-based approach is not adequate for VizRepos that contain a significant number of visualizations with little or no text. In that case, other techniques, such as collaborative filtering, could be more appropriate.

Although we have focused on Tableau workbook specifications, most visualization systems, such as Plotly, MS Power BI, or Looker, have similar textual features for describing data visualizations. While feature extraction would depend on the specification format, the basic processes we outline in this paper and the findings of the user study should be applicable to any repository of computer-readable viz specifications.




\subsection{Near-Duplicates Grouping}

The proposed viz similarity measure can be utilized in other scenarios beyond content-based recommendations per se. One such example is the grouping of near-duplicates (challenge C4 in ~\autoref{sec:data-challenges}) when browsing VizRepos. A keyword search typically results in hundreds of potentially useful results but is often cluttered because of masses of near-duplicates. To save precious screen real estate and avoid the need to inspect many repetitive workbooks, similar items could be grouped together~\cite{joshi2011autoGroupingEmails,gemmell2006cleanLivingNearDuplicates}. One approach would be to show only a single representative workbook, for example the last modified one, with a symbol to indicate that it is a group proxy. Users could optionally explore near-duplicates (or different versions) in a detail view, similar to the facets in the proposed VizCommender.

\subsection{Hybrid and Sequence-Aware Recommender Systems}\label{sec:advanced-recommendations}

For this project we chose to create recommendations based on a single reference workbook as an obvious first step, to serve as a base for and inspire future studies. In this scenario, 
the user visits a workbook detail page and gets recommendations that lead to additional insights. The query is comparable to the frequently used search-by-example paradigm that requires active user input. Content-based recommender systems are nowadays used in many scenarios as well, and often turn into more sophisticated hybrid systems~\cite{ccano2017hybridRecSys} by incorporating additional user data and behavioral patterns. For example, the similarity model can take the popularity of workbooks into account when ranking recommendations. Another direction for future work would be a sequence-aware model~\cite{quadrana2018sequenceAwareRecSys} which could record user sessions and base recommendations on a collection of relevant workbooks rather than a single query.
Further, we speculate that our approach might prove adaptable to other repositories based on annotated, semantically rich, text-based specifications such as spreadsheets, electronic circuits, and computer-aided design.
 
\section{Conclusion}

We present the concept of a text-based topic similarity measure to be used towards content-based recommendations in visualization repositories. We investigate four applicable NLP models and conduct a user study that demonstrates it is possible to obtain good alignment between human similarity perception and off-the-shelf model predictions, despite the special data characteristics and associated issues. An iterative process and continuous engagement with the Tableau RecSys Group resulted in VizCommender, a proof-of-concept system that uses an LDA-based model to provide recommendations based on a reference viz workbook. While our prototype is implemented for Tableau workbooks, we believe targeted recommendations can significantly facilitate information seeking in all kinds of visualization repositories that are defined by similar text-based specifications.



\bibliographystyle{abbrv-doi}

\bibliography{template}
\end{document}